\documentclass[aps,prd,showpacs,onecolumn,floatfix,nofootinbib,notitlepage,superscriptaddress,showkeys,longbibliography, amsmath,amssymb,amsfonts]{revtex4-1} 

\usepackage{bm}
\usepackage{slashed}
\usepackage{graphicx}

\usepackage{amsmath,amssymb,amsfonts,graphicx}
\usepackage{dsfont, mathrsfs}
\usepackage{bm}
\usepackage{slashed}
\usepackage{subfigure}
\usepackage[svgnames]{xcolor}
\usepackage{placeins}
\usepackage{array}
\usepackage{natbib,hyperref}  
\usepackage{url}
\usepackage{lineno}

\makeatletter
\newcommand{\thickhline}{%
    \noalign {\ifnum 0=`}\fi \hrule height 1pt
    \futurelet \reserved@a \@xhline
}
\newcolumntype{"}{@{\hskip\tabcolsep\vrule width 1pt\hskip\tabcolsep}}
\makeatother

\usepackage{xspace}

\newcommand{\Eq}[1]{Eq.~(\ref{#1})}
\newcommand{\al}[1]{\begin{align} #1 \end{align}}
\newcommand{\non}{\nonumber}
\newcommand{\vect}[1]{{\bf{#1}}}

\def\fh{\varphi_h}

\def\fR{\varphi_R}
\def\fRB{\varphi_{\bar{R}}}

\def \fk{\varphi_{k}}

\def\fKR{\varphi_{KR}}
\def\fKRB{\varphi_{\bar{K} \bar{R}}}
\def \fq{\varphi_{q}}
\def \fqR{\varphi_{qR}}
\def \fqRB{\varphi_{q\bar{R}}}

\def\kT{\vect{k}_T}
\def\kBT{\vect{\bar{k}}_T}
\def\RT{\vect{R}_T}

\def\RBT{\vect{\bar{R}}_T}
\def\qT{\vect{q}_T}

\def\pT{\vect{p}_T}

\newcommand{\Ge}{\mathrm{GeV}}

\newcommand{\ImS}{0.47\columnwidth}

\newcommand{\GapCapt}{\vspace{-8pt}}

\begin{document}

\title{Accessing quark helicity in $e^+e^-$ and SIDIS via dihadron correlations.} 

\preprint{ADP-18-18/T1066}

\author{Hrayr~H.~Matevosyan}
\thanks{ORCID: http://orcid.org/0000-0002-4074-7411}
\affiliation{ARC Centre of Excellence for Particle Physics at the Terascale,\\ 
and CSSM, Department of Physics, \\
The University of Adelaide, Adelaide SA 5005, Australia
\\ http://www.physics.adelaide.edu.au/cssm
}

\begin{abstract}
The correlation between the longitudinal polarization of a fragmenting quark and the transverse momenta of the produced hadrons was predicted over two decades ago. Nevertheless, experimental searches in the electron-positron annihilation process, both through the so-called jet handedness measurements by the {\tt SLD} Collaboration and more recently via the measurements of the azimuthal asymmetry containing the helicity-dependent dihadron fragmentation function (DiFF) by the {\tt BELLE} Collaboration, did not yield a signal. We will first discuss our recent explanation of the zero result at {\tt BELLE},  and the two new methods for accessing the helicity-dependent DiFFs both in the electron-positron annihilation experiments, and in the semi-inclusive deep inelastic scattering (SIDIS) experiments with a longitudinally polarized target. We will also for the first time describe yet another, new method for accessing the helicity-dependent DiFFs in SIDIS using polarized beam asymmetry. Finally, we will present a new Monte Carlo calculation of the specific Fourier moments of the helicity-dependent DiFF entering in to the new asymmetries, performed within the extended quark-jet model, and compare the results to those for the interference DiFF.
\end{abstract}


\date{\today}                                           

\maketitle

\label{SEC_INTRO}
 
 The idea of accessing the longitudinal polarization of a hadronizing quark in deep inelastic scattering experiments has been first proposed over 25 years ago by Efremov et al.~\cite{Efremov:1992pe} using the measurements of the so-called longitudinal jet handedness. More recently, another proposal by Boer et al.~\cite{Boer:2003ya} for the measurement of the correlations of the quark and antiquark longitudinal polarization in electron-positron annihilation involve azimuthal modulations of the cross section for inclusive production of two back-to-back hadron pairs. Here, the relevant term in the cross section involves convolutions of two quark helicity-dependent dihadron fragmentation functions (DiFF), conventionally denoted as $G_1^\perp$. It describes the correlations between the longitudinal polarization of a fragmenting quark with the vector product of the transverse momenta of a produced hadron pair, with respect to three-momentum of the quark. Nevertheless, the experimental searches for such effects by the {\tt BELLE} Collaboration failed to detect a signal~\cite{Vossen:2015znm}, which is in contrast to the earlier successful measurements of asymmetry for the quark transverse-polarization-dependent interference DiFF (IFF).
 
  These observations, together with the inconsistencies in the definitions of the IFFs entering into asymmetries for $e^+e^-$ annihilations and SIDIS  measurements, highlighted in Ref.~\cite{Matevosyan:2017uls}, led us to revisit the derivation of the cross section calculations of Ref.~\cite{Boer:2003ya}. Our findings, detailed in Ref.~\cite{Matevosyan:2018icf}, show that an apparent error in the kinematics definitions in Ref.~\cite{Boer:2003ya} have lead to a wrong expression for the corresponding cross section, and the derived azimuthal asymmetries. In particular, as we first noted in Ref.~\cite{Matevosyan:2017liq}, the proposed asymmetry for accessing $G_1^\perp$ should vanish. There, we also proposed two new measurements of asymmetries involving $G_1^\perp$, one in electron-positron annihilation and one in SIDIS processes.
  
 Let us first briefly describe the electron-positron annihilation process, where we consider two back-to-back inclusive hadron pair creation, $e^+e^-\to h_1 h_2 + \bar{h}_1 \bar{h}_2 + X$. In the leading order approximation, and for the center-of-mass energies well below the mass of the $Z^0$ boson, this process is described by the annihilation of the lepton pair into a virtual photon, that, in turn produces a quark and an antiquark pair, $e^+e^- \to \gamma^* \to a \bar{a}$. The quark and antiquark hadronize into back-to-back jets, producing the considered hadron pairs. We denote the momenta of the electron-positron pair as $l$ and $l'$, the momenta of the quark and antiquark as $k$ and $\bar{k}$, and the momentum of the virtual photon is $q = l+l' =  k+ \bar{k}$. The masses and the momenta of the hadron pair in the jet of the quark are taken to be $M_1, M_2$ and $P_1, P_2$, and those on the other side produced in the jet of the antiquary as  $\bar{M}_1, \bar{M}_2$ and $\bar{P}_1, \bar{P}_2$. In dihadron studies, the individual momenta of the pairs are substituted with their total and relative momenta %
\al
{
 P_h &\equiv P_1 + P_2,
 \\
 R &\equiv \frac{1}{2}\big( P_1 -P_2\big),
}
and the invariant mass of the pair is denoted as $M_h^2 = P_h^2$.

We use the reference frame as described in detail in Sec.~III of Ref.~\cite{Matevosyan:2018icf}, where the $\hat{z}$ axis is taken to point opposite to the total 3-momentum of $\bar{h}_1,\bar{h}_2$ pair, while the $\hat{x}$ axis is taken along the transverse component of the electron's momentum to the $\hat{z}$ axis, as depicted in Fig.~2 of~\cite{Matevosyan:2018icf}. The cross section of this process can be then derived by factorizing the production of the hadron pairs on each side using the corresponding quark-quark correlators, where we choose $Q^2 = q^2$ as the hard scale in the process. Within the leading hadron approximation, where we assume  $P_h \cdot \bar{P}_h \sim Q^2$, and ignoring all the corrections of order $1/Q$,  the relevant contributions from the unpolarized and helicity-dependent DiFFs are given by
\al
{
\label{EQ_XSEC}
\frac{d \sigma_{U,L}^{e^+e^- \to (h_1 h_2) (\bar{h}_1 \bar{h}_2) X }}{d^2 \qT \ d\fR\  d\fRB
\ d^7 V 
} 
 = A(y)\sum_{a,\bar{a}}\frac{3 \alpha^2  e_a^2 }{\pi Q^2}  &
\Big\{
\mathcal{F}\Big[ D_1^a \bar{D}_1^{\bar{a}}\Big]
\\\non
&
- \mathcal{F}\Big[
-\frac{|\RT| |\kT| }{M_h^2} \frac{|\RBT|  |\kBT| }{\bar{M}_h^2} \sin(\fKR) \sin(\fKRB) G_1^{\perp a} \bar{G}_1^{\perp \bar{a}}  \Big]
\Big\}
 ,
}
where the sum is over all possible flavors $a$ of produced quarks with  charges $e_a$, and $\alpha$ is the electromagnetic coupling constant. The light cone-momentum fractions are given by $z = P_h^+/k^+$ and $\bar{z} = \bar{P}_h^-/\bar{k}^-$, and the transverse momentum of the virtual photon is $\qT = - \vect{P}_h/z$. The azimuthal angles of $\RT$ and $\RBT$ are denoted as  $\fR$ and $\fRB$, while $\fKR \equiv  \fk -\fR$ is the relative azimuthal angle between $\kT$ and $\RT$. The rest of the phase space factors are denoted by $d^7V$, and will not be central in the further arguments. The unpolarized and the helicity-dependent DiFFs $D_1^a$ and $G_1^{\perp a}$ are functions of total $z$ and relative $\xi = 1/2 + R^+/P_h^+$ light-cone momentum fractions, as well as  $\kT$ and $\RT$. The transverse momenta of the quark and the antiquark in the cross section are convoluted due to the momentum conservation
\al
{
\label{EQ_F_CONVOL}
\mathcal{F}[w D^a \bar{D}^{\bar{a}} ]
 =  \int d^2 \kT &d^2 \kBT 
 \  \delta^2(\vect{k}_T + \vect{\bar{k}}_T - \vect{q}_T)
  w( \kT, \kBT, \RT, \RBT)\ D^a \ D^{\bar{a}}.
}

The only azimuthal dependence in the DiFFs appear as functions of the relative azimuthal angle $\fKR$ ( $\fKRB$ for the antiquark ones), that allows us to write their Fourier cosine decomposition
\al
{
\label{EQ_D_FOURIER}
D^a(z, \xi, \kT^2, \RT^2,& \cos(\fKR)) 
  = \frac{1}{\pi} \sum_{n=0}^\infty
 \frac{\cos(n \cdot \fKR)}{1+\delta^{0,n}} \ D^{a,[n]}(z, \xi, \kT^2, \RT^2),
}
which will be useful in calculations of various weighted integrals of the cross section. For example, the expression for the  cross section  integrated  over $\qT, \fR, \fRB$  without any additional factors
\al
{
\label{EQ_XSEC_INT_UPOL}
 \langle 1 \rangle 
 &=  \int d \sigma_{U,L}^{e^+e^- \to (h_1 h_2) (\bar{h}_1 \bar{h}_2) X }\  \times 1
 =   \frac{3 \alpha^2}{\pi Q^2}  A(y)
 \sum_{a,\bar{a}} e_a^2  \ D_1^a(z, M_h^2) \bar{D}_1^{\bar{a}}(\bar{z}, \bar{M}_h^2),
}
involves only the zeroth Fourier cosine moment of the unpolarized DiFF
\al
{
D_1^{a}(z, M_h^2)  =  z^2 \int d^2 \kT \int  d \xi
\ D_1^{a,[0]}(z, \xi, \kT^2, \RT^2) \, .
}

 We can access the helicity-dependent DiFFs by measuring weighted integrated cross section. It is clear, that for a weight factor that contains no dependence on $\qT$, the integration over $\qT$ in \Eq{EQ_XSEC} will completely disentangle the quark and antiquark momentum convolutions of \Eq{EQ_F_CONVOL}. Then, the $G_1^\perp$  terms will vanish after integration over $\fk$, since a trivial change of variables $\fk \to \fKR$ will yield an integration of a $\sin(\fKR)$ multiplied by cosine moments of $\fKR$, tha trivially vanish due to the orthogonality relations. Thus the asymmetry derived in~\cite{Boer:2003ya} for measuring the first Fourier cosine moment of $G_1^\perp$
\al
{
\langle \cos(2(\fR - \fRB)) \rangle =0,
}
should also vanish, as  confimred in the recent measurements by the {\tt BELLE} Collaboration~\cite{Vossen:2015znm}. 

 A new asymmetry to measure $G_1^\perp$ has been recently  proposed in Ref~\cite{Matevosyan:2017liq}, where the weight includes both the azimuthal angle and the magnitude of $\qT$
\al
{
\label{EQ_PHI1_AV}
\left\langle \frac{q_T^2 \Big(3 \sin(\fqR) \sin(\fqRB)
+ \cos(\fqR) \cos(\fqRB) \Big)}{M_h \bar{M}_h  }
 \right \rangle&
= \frac{12 \alpha^2 A(y) }{\pi Q^2} 
 \sum_{a,\bar{a}}  e_a^2  
 G_1^{\perp a}(z, M_h^2) \bar{G}_1^{\perp \bar{a}}(\bar{z}, \bar{M}_h^2),
}
where $\fqR\equiv \fq- \fR$, and the integrated $G_1^\perp$ is now defined as the combination
\al
{
\label{EQ_G1_INT}
G_1^{\perp a}(z, M_h^2) \equiv  G_1^{\perp a,[0]}(z, M_h^2)  - G_1^{\perp a,[2]}(z, M_h^2),
}
of the Fourier cosine moments of $G_1^\perp$, which are defined as
\al
{
\label{EQ_G_MOMS}
G_1^{\perp a,[n]}(z, M_h^2) \equiv  &\ z^2  \int d^2 \kT  \int  d \xi 
 \left(\frac{\kT^2}{2 M_h^2}\right)  \frac{|\RT|}{M_h} \ G_1^{\perp a,[n]}(z, \xi, \kT^2, \RT^2).
}
The combination of the two terms in \Eq{EQ_PHI1_AV} is chosen such as to cancel the contributions from the first cosine moments of the unpolarized DiFFs. It is useful to also write the expression for the corresponding azimuthal asymmetry
\al
{
\label{EQ_ee_SSA}
 A_{e^+e^-}^{\Rightarrow}(z ,& \bar{z}, M_h^2, \bar{M}_h^2)
 = 4
\frac{ \sum_{a,\bar{a}} e_a^2\
 G_1^{\perp a}(z, M_h^2) \ G_1^{\perp \bar{a}}(\bar{z}, \bar{M}_{h}^2)
 }
 { \sum_{a,\bar{a}} e_a^2\
 D_1^a(z, M_h^2) \ D_1^{\bar{a}} (\bar{z}, \bar{M}_{h}^2).
 },
}

 It is also interesting to consider the azimuthal asymmetries induced by $G_1^\perp$ in dihadron production in SIDIS. The reason is that being a chiral-even function, $G_1^\perp$ can couple to well measured chiral even PDFs. This is in contrast to the IFF, which is used to extract the chiral-odd transversity PDF from SIDIS measurements. The process considered here is $l + N \to l' + h_1h_2 +X$, as detailed in Ref.~\cite{Matevosyan:2017liq}, where at leading order a virtual photon emitted by the initial (polarized) lepton $l$ knocks out a quark from the polarized  target nucleon $N$, which then hadronizes, producing $h1,h2$, etc. The corresponding kinematics and the cross section derivations are described in Ref.~\cite{Bacchetta:2002ux}. Here, we will consider the unpolarized part of the cross section $\sigma_{UU}$, and the term $\sigma_{UL}$ involving $G_1^\perp$ for a target with longitudinally polarization $S_L$ 
\al
{
\label{EQ_SIDIS_XSEC_UU}
\frac{d \sigma_{UU} + d \sigma_{UL} }{ d^2 \vect{P}_{h\perp} d\fR  \ d^6V'}
=  \frac{\alpha^2 }{\pi y Q^2} A'(y)\sum_a  \ e_a^2\
\left\{ \mathcal{G} \Big[ f_{1}^a\ D_1^{ a}
 \Big]
 -  S_L 
 \mathcal{G} \left[\frac{|\RT| |\kT|}{M_h^2}\sin(\fKR) g_{1L}^a  G_1^{\perp a} \right]
 \right\} ,
}
where $\vect{P}_{h\perp}$ is again the transverse component of the pair's total momentum, $f_1$ and $g_{1L}$ are the unintegrated unpolarized and helicity PDFs, and $d^6V'$ denotes the remaining terms of the phase space element. The momentum conservation at quark-virtual photon vertex yields the convolution between the transverse momentum of the initial quark $\pT$ and the fragmenting quark $\kT$
\al
{
\non
\mathcal{G}[&w f^q D^{q} ] \equiv \int d^2 \pT  \int d^2 \kT
   \delta^2\Big(\kT -\pT+ \frac{\vect{P}_{h\perp}}{z} \Big)
    w( \pT, \kT, \RT)
f^q(x, p_T) D^{q}(z, \xi, k_T^2, R_T^2, \kT \cdot \RT  ).
}

First, we calculate the integrated cross section without any additional weighting, which breaks up the momentum convolution and yields a product of the  integrated unpolarized PDFs and DiFFs
\al
{
& \left\langle 1 \right\rangle 
 = \int d \sigma_{UU} \times 1
=  \sum_a \frac{2 \alpha^2 e_a^2}{y\ Q^2} A'(y) f_1^a(x) z^2 D_1^a (z, M_h^2).
}

 We can again construct an appropriate weighting factor to access the helicity-dependent DiFF
\al
{
 \left\langle \frac{P_{h\perp} \sin(\fh - \fR)}{M_h} \right\rangle 
=   S_L \ A'(y)\sum_a \frac{2 \alpha^2 e_a^2}{y\ Q^2} 
\  g_{1L}^a(x) \ z\ G_1^{\perp a}(z, M_h^2),
}
where the integrated DiFF $G_1^{\perp}$ is exactly the same as in \Eq{EQ_G1_INT}. The corresponding  asymmetry is
\al
{
\label{EQ_SIDIS_SSA}
 A^{\Rightarrow}_{UL}(x, z , M_h^2)
 =&S_L \frac{\sum_{a} e_a^2\  g_{1L}^a(x)
 \ z\ G_1^{\perp a}(z, M_h^2) }
 { \sum_{a} e_a^2\ 
f_1^a(x)\ D_1^a(z, M_h^2)  }.
}

Moreover, we can also measure $G_1^\perp$ by considering a polarized lepton beam asymmetries for an unpolarized target. The relevant term w in Ref.~\cite{Bacchetta:2002ux} is similar in structure to $\sigma_{UL}$

\al
{
 \label{EQ_XSEC_LU}
 \frac{d \sigma_{LU}} {d^2 \vect{P}_{h\perp} d\fR \ d^6V'}
 = -  \lambda_l   \frac{\alpha^2 }{\pi y Q^2} C'(y)
  \sum_a  e_a^2 \
  &  \mathcal{G} \Big[
   	\frac{|\kT| |\RT|}{M_h^2}  \sin(\fKR) \ f_{1}^a\ G_1^{\perp a} \Big]
 ,
}
where $\lambda_l$ is the helicity of the incoming lepton.
The corresponding single spin asymmetry
\al
{
 A_{LU}^{\Rightarrow}(x,y,z,M_h^2) = \frac{1}{M_h }\dfrac{ \left\langle P_{h\perp} \sin(\fh - \fR) \right\rangle }{ \ \left\langle 1 \right\rangle } =   \lambda_l \frac{C'(y)}{A'(y) } \frac{\sum_a e_a^2 \ f_1^a(x) \  z\ G_1^{\perp a}(z, M_h^2)}{\sum_a e_a^2 \ f_1^a(x) \ D_1^{ a}(z, M_h^2)},
}
now involves the unpolarized PDF $f_1$ in the numerator instead of the helicity PDF in \Eq{EQ_SIDIS_SSA}, which should enhance the signal for comparable values of the longitudinal polarizations of the corresponding initial state particles. The penalty here is the additional factor $C'(y)/A'(y)$.

\begin{figure}[tb]
\centering 
\includegraphics[width=\ImS]{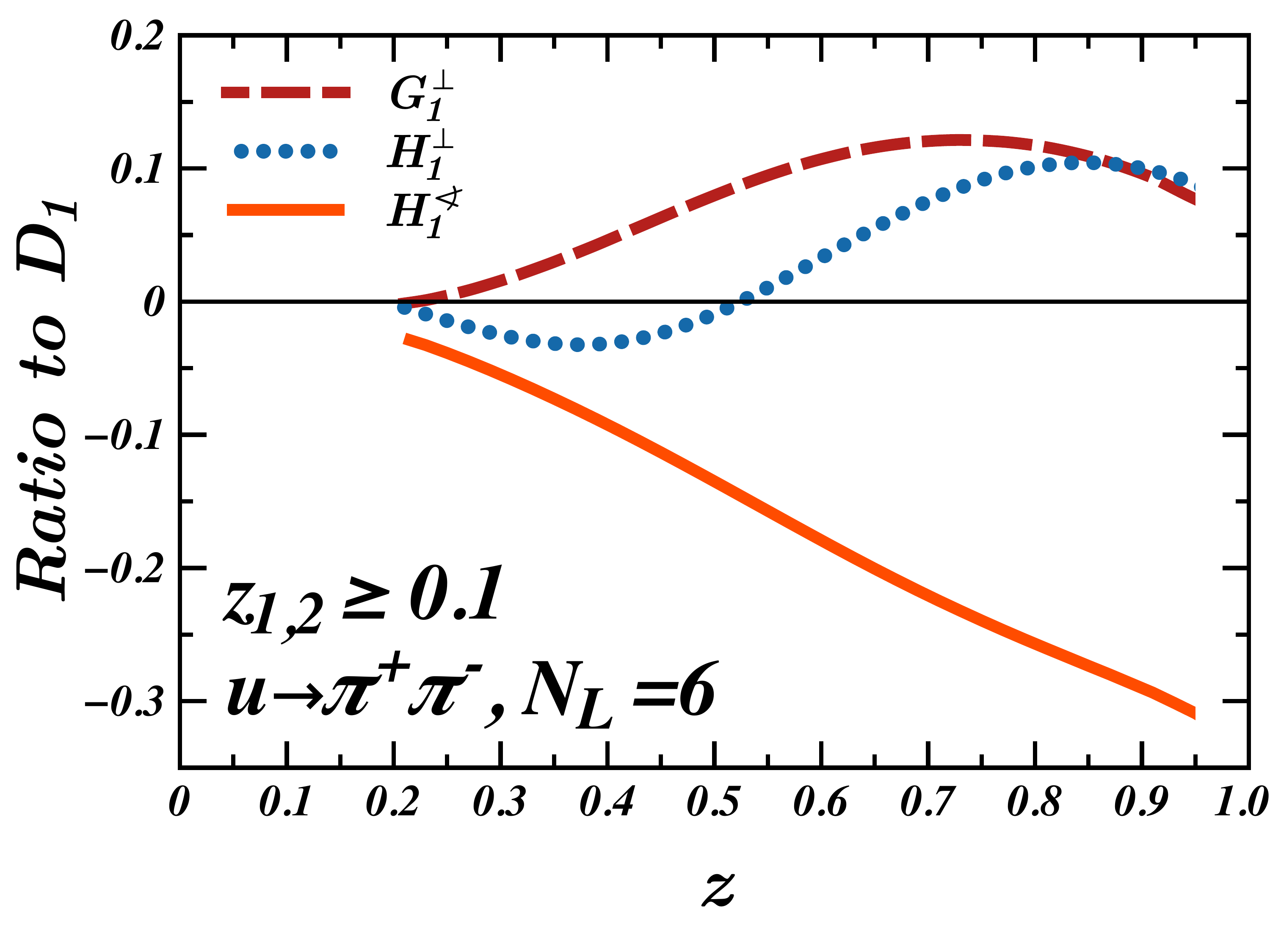}
\GapCapt
\vspace{-0.2cm}
\caption{The extended quark-jet model results for the analyzing powers of all three integrated polarized DiFFs for $u\to \pi^+\pi^-$ pairs. A minimum cut on light-cone momentum fractions of both hadrons $z_{1,2}\geq 0.1$ is imposed, and the number of produced hadrons in the quark hadronization chain is limited to $N_L=6$.}
\label{PLOT_QUARK_JET}
\end{figure}

We recently used the extended quark-jet model~\cite{Bentz:2016rav,*Matevosyan:2016fwi} to calculate all three polarized DiFFs~\cite{Matevosyan:2017alv,Matevosyan:2017uls}. Several of the lowest Fourier cosine moments of $G_1^\perp$ were calculated in Ref.~\cite{Matevosyan:2017alv}, but the final results were presented only for the first cosine moment entering into the asymmetry proposed in~\cite{Boer:2003ya}. In Fig.~\ref{PLOT_QUARK_JET} we present an updated quark-jet calculation of the analyzing power for the new definition of $G_1^\perp$, as well as for the two transverse polarization dependent DiFFs for comparison.
 
In conclusion, we discussed the recently derived asymmetries for accessing the helicity-dependent DiFFs $G_1^\perp$ both in electron-positron annihilation and SIDIS production of dihadron pairs, that were first presented in Ref.~\cite{Matevosyan:2017liq}. For the former process, we first elaborated why the previously proposed asymmetries should vanish, using the newly re-derived cross section for this process~\cite{Matevosyan:2018icf}. We also described the newly proposed asymmetry, that gives access to the combination of the zeroth and second Fourier cosine moments of $G_1^\perp$. Here, the weighting with the square of the component of the total 3-momentum of one of the pairs transverse to that of the other pair on the opposite side allows us to break-up the convolution of two DiFFs from each side into a simple product. For SIDIS measurements, we presented the target longitudinal spin asymmetry that measures a product of the well determined helicity PDF and the same integrated  $G_1^\perp$ that enters in $e^+e^-$ asymmetry. Further, we proposed a new beam longitudinal single spin asymmetry for SIDIS experiments, that measures now a product of the unpolarized PDF and the integrated $G_1^\perp$. The two complementary SIDIS measurements, that can be performed both at JLab~$12\Ge$ and the planned Electron-Ion-Collider, should enable us to extract the helicity-dependent DiFFs by leveraging our knowledge of the well determined unpolarized and helicity PDFs. Similar measurements of $e^+e^-$ can be performed at the upcoming {\tt BELLE~II} experiment, which then can be used to directly test the universality of $G_1^\perp$ between SIDIS and $e^+e^-$. Finally, in Fig.~\ref{PLOT_QUARK_JET} we presented our quark-jet model calculations of the relative size of $G_1^\perp$ compared to the IFF , showing that in the region of moderate $z$ values, the magnitude of the helicity-dependent DiFF is about half of that for IFF. Thus, the future experimental measurements are promising, as the IFF asymmetries have been measured both in SIDIS and $e^+e^-$.

This work was supported by the Australian Research Council through the ARC Centre of Excellence for Particle Physics at the Terascale (CE110001104), and by the ARC  Discovery Project No. DP151103101, as well as by the University of Adelaide. 

\bibliographystyle{JHEP}
\bibliography{fragment}

\providecommand{\href}[2]{#2}\begingroup\raggedright\begin{thebibliography}{10}

\bibitem{Efremov:1992pe}
A.~Efremov, L.~Mankiewicz and N.~Tornqvist, \emph{{Jet handedness as a measure
  of quark and gluon polarization}},
  \href{https://doi.org/10.1016/0370-2693(92)90451-9}{\emph{Phys.Lett.}
  {\bfseries B284} (1992) 394}.

\bibitem{Boer:2003ya}
D.~Boer, R.~Jakob and M.~Radici, \emph{{Interference fragmentation functions in
  electron positron annihilation}},
  \href{https://doi.org/10.1103/PhysRevD.67.094003}{\emph{Phys.Rev.} {\bfseries
  D67} (2003) 094003} [\href{https://arxiv.org/abs/hep-ph/0302232}{{\ttfamily
  hep-ph/0302232}}].

\bibitem{Vossen:2015znm}
A.~Vossen, \emph{{Recent Fragmentation Function Measurements at Belle}},
  {\emph{PoS} {\bfseries DIS2015} (2015) 216}.

\bibitem{Matevosyan:2017uls}
H.~H. Matevosyan, A.~Kotzinian and A.~W. Thomas, \emph{{Dihadron fragmentation
  functions in the quark-jet model: Transversely polarized quarks}},
  \href{https://doi.org/10.1103/PhysRevD.97.014019}{\emph{Phys. Rev.}
  {\bfseries D97} (2018) 014019}
  [\href{https://arxiv.org/abs/1709.08643}{{\ttfamily 1709.08643}}].

\bibitem{Matevosyan:2018icf}
H.~H. Matevosyan, A.~Bacchetta, D.~Boer, A.~Courtoy, A.~Kotzinian, M.~Radici
  et~al., \emph{{Semi-inclusive production of two back-to-back hadron pairs in
  $e^+e^-$ annihilation revisited}},
  \href{https://doi.org/10.1103/PhysRevD.97.074019}{\emph{Phys. Rev.}
  {\bfseries D97} (2018) 074019}
  [\href{https://arxiv.org/abs/1802.01578}{{\ttfamily 1802.01578}}].

\bibitem{Matevosyan:2017liq}
H.~H. Matevosyan, A.~Kotzinian and A.~W. Thomas, \emph{{Accessing quark
  helicity through dihadron studies}},
  \href{https://doi.org/10.1103/PhysRevLett.120.252001}{\emph{Phys. Rev. Lett.}
  {\bfseries 120} (2018) 252001}
  [\href{https://arxiv.org/abs/1712.06384}{{\ttfamily 1712.06384}}].

\bibitem{Bacchetta:2002ux}
A.~Bacchetta and M.~Radici, \emph{{Partial wave analysis of two hadron
  fragmentation functions}},
  \href{https://doi.org/10.1103/PhysRevD.67.094002}{\emph{Phys. Rev.}
  {\bfseries D67} (2003) 094002}
  [\href{https://arxiv.org/abs/hep-ph/0212300}{{\ttfamily hep-ph/0212300}}].

\bibitem{Bentz:2016rav}
W.~Bentz, A.~Kotzinian, H.~H. Matevosyan, Y.~Ninomiya, A.~W. Thomas and
  K.~Yazaki, \emph{{Quark-Jet model for transverse momentum dependent
  fragmentation functions}},
  \href{https://doi.org/10.1103/PhysRevD.94.034004}{\emph{Phys. Rev.}
  {\bfseries D94} (2016) 034004}
  [\href{https://arxiv.org/abs/1603.08333}{{\ttfamily 1603.08333}}].

\bibitem{Matevosyan:2016fwi}
H.~H. Matevosyan, A.~Kotzinian and A.~W. Thomas, \emph{{Monte Carlo
  Implementation of Polarized Hadronization}},
  \href{https://doi.org/10.1103/PhysRevD.95.014021}{\emph{Phys. Rev.}
  {\bfseries D95} (2017) 014021}
  [\href{https://arxiv.org/abs/1610.05624}{{\ttfamily 1610.05624}}].

\bibitem{Matevosyan:2017alv}
H.~H. Matevosyan, A.~Kotzinian and A.~W. Thomas, \emph{{Dihadron fragmentation
  functions in the quark-jet model: Longitudinally polarized quarks}},
  \href{https://doi.org/10.1103/PhysRevD.96.074010}{\emph{Phys. Rev.}
  {\bfseries D96} (2017) 074010}
  [\href{https://arxiv.org/abs/1707.04999}{{\ttfamily 1707.04999}}].

\end{thebibliography}\endgroup

\end{document}